\documentstyle[aps,twocolumn]{revtex}

\begin{document}
\draft
\title{
A position-momentum EPR state of distantly-separated trapped
atoms
}

\author{A.S. Parkins}
\address{
Department of Physics, University of Auckland, Auckland, 
New Zealand
}
\author{H.J. Kimble}
\address{
Norman Bridge Laboratory of Physics 12-33, California 
Institute of Technology, Pasadena, CA 91125, U.S.A.
}

\date{\today}

\maketitle
\begin{abstract}
We propose a scheme for preparing an EPR state in position and
momentum of a pair of distantly-separated trapped atoms.
The scheme utilizes the entangled light fields output from
a nondegenerate optical parametric amplifier. 
Quantum state exchange between these fields and the motional
states of the trapped atoms is accomplished via interactions
in cavity QED.
\end{abstract}



\section{
Introduction
}

In 1935, Einstein, Podolsky and Rosen (EPR) \cite{Einstein35} 
proposed a now famous gedanken experiment involving a system of 
two particles spatially separated but correlated in position and
momentum as described by the Wigner function
\begin{equation} \label{EPRstate}
W(q_{1},p_{1};q_{2},p_{2}) \propto 
\delta (q_{1}+q_{2})\delta (p_{1}-p_{2}),
\end{equation}
where $q_1$ and $q_2$ are the continuous position variables of
the particles with corresponding conjugate momenta $p_1$ and
$p_2$.
With the assumption of local realism, but with the 
{\em apparent} ability to assign definite values to canonically 
conjugate variables of one particle from measurements of the 
other particle in this system, a conflict with the Heisenberg
uncertainty principle seemingly follows, which led EPR to 
conclude that quantum mechanics is incomplete.
Bohm \cite{Bohm51} adapted this argument to a system of 
discrete (dichotomic) variables, to which Bell applied his 
classic analysis, deriving the so-called Bell inequalities 
\cite{Bell65,Bell87,Clauser78} which quantify explicitly the 
conflict between local realism and quantum mechanics. 
Note that although measurements of $(q_i,p_i)$ do not lead
to a violation of a Bell inequality for the original EPR
state of Eq.~(\ref{EPRstate}) (Ref.\cite{Bell87}, p.196), the
entanglement of this state guarantees that an appropriate set
of variables exists for which a contradiction with local
realism would be manifest \cite{Peres93,Banaszek98}.

Experimental demonstrations of the conflict between
quantum mechanics and local realism
have concentrated almost exclusively on systems of discrete
variables, such as electron spin or photon polarization
\cite{Clauser78,Expts}.
Only one experiment, by Ou {\em et al}. \cite{Ou92a,Ou92b}, 
following suggestions by Reid and Drummond \cite{Reid88},
has in fact realized the EPR paradox as originally envisioned 
by EPR; that is, for canonically conjugate variables with a 
continuous spectrum. The EPR ``source'' in this experiment was 
a nondegenerate optical parametric amplifier (NOPA),
and the relevant variables were the quadrature amplitudes 
of the entangled electromagnetic fields generated in 
the parametric process. These amplitudes are analogous to the 
position and momentum of a particle and can be measured very 
efficiently via homodyne detection \cite{Kimble92}.

In the present work, we describe a new scheme that goes beyond 
an analogy and actually realizes an EPR state in position and
momentum for a pair of massive particles at distinct physical
locations. 
Our proposal for achieving {\em stored} entanglement for
continuous quantum variables is based upon a set of
interactions in cavity quantum electrodynamics (QED) that
allows for the exchange of quantum states between the motion
of trapped atoms and propagating light fields \cite{Parkins99}. 
By exploiting these interactions and the light source of 
\cite{Ou92a,Ou92b}, we show that it should be possible to prepare
{\em deterministically} a state of the form (\ref{EPRstate})
for a pair of trapped atoms located at macroscopically-separated
sites. 
Beyond conventional $(q,p)$ projections as in homodyne or
heterodyne measurements, the setting of atom traps and cavity
QED also enables detection strategies for the explicit 
demonstration of the nonlocal character of the resulting
EPR state. Moreover, the techniques that we describe could be
important resources for the realization of quantum networks,
a particular example being the creation of EPR states
to enable the teleportation of the center-of-mass wave
function of a massive particle \cite{Parkins20}.

\section{
Trapped atom coupled to an optical cavity mode
}

We begin with the basic setup that facilitates the motion-light
coupling fundamental to our scheme \cite{Parkins99}; this
setup was originally considered by Zeng and Lin \cite{Zeng94}.
We consider a single two-level atom (or ion) confined
in a harmonic trap located inside an optical 
cavity. The atomic transition of frequency
$\omega_{\rm a}$ is coupled to a single mode of the cavity field of
frequency $\omega_{\rm c}$ and is also assumed to be driven by an
external (classical) laser field of frequency $\omega_{\rm L}$.
The physical setup and excitation scheme are
depicted in Fig.~1. The cavity is aligned along the $x$-axis, while
the laser field is incident from a direction in the $y$-$z$ plane
(i.e., perpendicular to the $x$-axis).

The Hamiltonian describing the internal and external atomic degrees
of freedom plus the atom-cavity and atom-laser couplings takes the  
form (in a frame rotating at the laser frequency for the 
internal-atomic and cavity operators)
\begin{eqnarray}
\hat{H}_0 = && 
\sum_{j=x,y,z} 
\hbar\nu_j (\hat{b}_j^\dagger\hat{b}_j+1/2) + \hbar\delta
\hat{a}^\dagger\hat{a} + \hbar\Delta\hat{\sigma}_+\hat{\sigma}_-
\nonumber
\\
&& \;\;\; +\, \hbar 
\left[ {\cal E}_{\rm L}(\hat{y},\hat{z},t) 
\hat{\sigma}_+ + 
{\cal E}_{\rm L}^\ast (\hat{y},\hat{z},t)
\hat{\sigma}_- \right] \nonumber
\\
&& \;\;\; +\, \hbar
g_0 \sin (k\hat{x}) (\hat{a}^\dagger
\hat{\sigma}_- + \hat{\sigma}_+\hat{a} ) .
\end{eqnarray}
Here, $\{\nu_x,\nu_y,\nu_z\}$ are the harmonic oscillation 
frequencies along the principal axes of the trap, 
$\hat{b}_j$ and $\hat{a}$ are annihilation operators for the
quantized atomic motion and cavity field, respectively, 
$\hat{\sigma}_-=|g\rangle\langle e|$ is the atomic lowering 
operator, and $\delta =\omega_{\rm c}-\omega_{\rm L}$ and 
$\Delta =\omega_{\rm a}-\omega_{\rm L}$. 
The quantity ${\cal E}_{\rm L}(\hat{y},\hat{z},t)$ is the (possibly
time-dependent) amplitude of the laser field.
The single-photon atom-cavity dipole coupling strength is
given by $g_0$, while the sine function describes the standing wave
structure of the cavity field (we assume that the centre of the 
trap is located at a {\em node} of the cavity field), with 
$k=2\pi /\lambda$ the wavenumber of the field and 
$\hat{x}=[\hbar /(2m\nu_x)]^{1/2}(\hat{b}_x+\hat{b}_x^\dagger )$. 

In \cite{Parkins99} a number of assumptions and approximations are
made in order to simplify the model. In particular:
\begin{list}{$\bullet$}{}
\item
The detunings of the light fields from the atomic transition 
frequency are assumed to be very large 
(i.e., $\Delta\gg |{\cal E}_{\rm L}|,g_0,\delta,\nu_j$), enabling 
atomic spontaneous emission to be neglected and the internal atomic
dynamics to be adiabatically eliminated.
\item
Any forms of motional decoherence associated with the trap itself
are ignored.
\item
The size of the harmonic trap is assumed to be small compared to the
optical wavelength (Lamb-Dicke regime), enabling the approximations
$\sin (k\hat{x})\simeq \eta_x (\hat{b}_x+\hat{b}_x^\dagger )$, where
$\eta_x$ ($\ll 1$) is the Lamb-Dicke parameter, and 
${\cal E}_{\rm L}(\hat{y},\hat{z},t)\simeq
{\cal E}_{\rm L}(t)e^{-i\phi_{\rm L}}$.
\item
The cavity and laser fields are tuned so that 
$\delta =\omega_{\rm c}-\omega_{\rm L}=\nu_x$.
\item
The trap frequency $\nu_x$ and cavity field decay rate $\kappa_a$
are assumed to satisfy 
$\nu_x\gg\kappa_a\gg |(g_0\eta_x/\Delta ){\cal E}_{\rm L}(t)|$.
The first inequality allows a rotating-wave approximation to
be made with respect to the trap oscillation frequency, while
the second inequality enables an adiabatic elimination of
the cavity field mode.
\end{list}
Given these conditions one can show that the motional mode 
dynamics in the $x$ direction can be described by the simple 
quantum Langevin equation
\begin{equation}
\dot{\tilde{b}}_x \simeq 
- \Gamma (t) \tilde{b}_x + \sqrt{2\Gamma (t)} \,
\tilde{a}_{\rm in}(t)\; ,
\end{equation}
where $\tilde{b}_x=e^{i\nu_xt}\hat{b}_x$, 
$\Gamma (t)=[g_0\eta_x{\cal E}_{\rm L}(t)/\Delta ]^2/\kappa_a$,
and $\tilde{a}_{\rm in}(t)$, which satisfies the commutation
relation 
$[\tilde{a}_{\rm in}(t),\tilde{a}_{\rm in}^\dagger (t^\prime )]
=\delta (t-t^\prime )$, is the quantum noise operator 
describing the {\em input to the cavity field} (in a frame 
rotating at the cavity frequency).
In this way, the statistics of the input light field can be
``written onto'' the state of the oscillator.
In \cite{Parkins99} it was shown how this effect can be used
to efficiently prepare a squeezed state of the motion of the
trapped atom. 
Here we extend that work further to the generation
of entanglement between the motional states of trapped atoms at
separated sites. In particular, our protocol transfers 
entanglement from a pair of quantum-correlated light fields to 
a pair of trapped atoms in a process of {\em qu}antum 
{\em s}tate {\em ex}change, or {\em qusex}.

\section{
Light source: nondegenerate parametric amplifier
}

Our source of quantum-correlated light fields is taken to be
a NOPA operating 
below threshold \cite{Ou92a,Ou92b,Kimble92}. The light fields  
may be nondegenerate in polarization or in frequency.
We denote the annihilation operators for the two intracavity
field modes, of frequencies $\omega_1$ and $\omega_2$, by 
$\hat{c}_1$ and $\hat{c}_2$, and the Hamiltonian
describing the coupling between these modes takes the form
(in a rotating frame)
\begin{equation}
H_{\rm I} = i\hbar\epsilon \left( \hat{c}_1\hat{c}_2 
- \hat{c}_1^\dagger \hat{c}_2^\dagger \right) \, ,
\end{equation}
where $\epsilon$ is the coupling strength, proportional to 
the nonlinear susceptibility of the intracavity medium and to 
the strength of the coherent pump field (at frequency 
$\omega_1+\omega_2$). 

Assuming the cavity mode amplitudes to be damped at the same
rate $\kappa_c$, equations of motion for the mode operators
(in the rotating frame) can be derived as 
\begin{equation}
\dot{\hat{c}}_{1,2} = -\kappa_c\hat{c}_{1,2} - 
\epsilon\hat{c}_{2,1}^\dagger + \sqrt{2\kappa_c}\, 
\hat{c}_{\rm in}^{(1,2)}(t) \, ,
\end{equation}
where $\hat{c}_{\rm in}^{(1,2)}(t)$ are the vacuum input 
fields to the NOPA cavity modes 
(see, e.g., \cite{Gardiner91,Walls94}).
The output fields from the NOPA then follow from the boundary
conditions
\begin{equation}
\hat{c}_{\rm out}^{(1,2)}(t) + \hat{c}_{\rm in}^{(1,2)}(t) =
\sqrt{2\kappa_c}\, \hat{c}_{1,2}(t) \, ,
\end{equation}
The (linear) equations above are readily solved in a 
Fourier-transformed space defined by 
$Z(\omega )=(2\pi )^{-1/2} \int dt\, Z(t)e^{-i\omega t}$.
Defining quadrature phase amplitudes (``positions'' and
``momenta'') for the output fields by
\begin{eqnarray}
X_{\rm out}^{(1,2)}(t) &=& \hat{c}_{\rm out}^{(1,2)}(t) +
\hat{c}_{\rm out}^{(1,2)\dagger}(t) \, , 
\\
Y_{\rm out}^{(1,2)}(t) &=& -i\left\{ 
\hat{c}_{\rm out}^{(1,2)}(t) - \hat{c}_{\rm out}^{(1,2)\dagger}(t) 
\right\} \, , 
\end{eqnarray}
the sum of the $X$ amplitudes is derived as \cite{Kimble92}
\begin{eqnarray}
X_{\rm out}^{(1)}(\omega ) && + X_{\rm out}^{(2)}(\omega )
\nonumber
\\
&& = \frac{\kappa_c-\epsilon +i\omega}{\kappa_c+\epsilon -i\omega}
\left\{ 
X_{\rm in}^{(1)}(\omega ) + X_{\rm in}^{(2)}(\omega )
\right\}
\nonumber
\\
&& \rightarrow 0 \;\;\; {\rm as}\;\; \epsilon\rightarrow\kappa_c
\;\; {\rm and}\;\; \omega\rightarrow 0 \, ,
\end{eqnarray}
while the difference of the $Y$ amplitudes is
\begin{eqnarray}
Y_{\rm out}^{(1)}(\omega ) && - Y_{\rm out}^{(2)}(\omega )
\nonumber
\\
&& = \frac{\kappa_c-\epsilon +i\omega}{\kappa_c+\epsilon -i\omega}
\left\{ 
Y_{\rm in}^{(1)}(\omega ) - Y_{\rm in}^{(2)}(\omega )
\right\}
\nonumber
\\
&& \rightarrow 0 \;\;\; {\rm as}\;\; \epsilon\rightarrow\kappa_c
\;\; {\rm and}\;\; \omega\rightarrow 0 \, .
\end{eqnarray}
So, the two output fields are highly correlated and, close to
$\omega =0$ and for $\epsilon\rightarrow\kappa_c$, their 
quadrature amplitudes exhibit precisely the properties of
the original EPR state, as demonstrated explicitly by the
Wigner function for the state of the (1,2) fields
\cite{Ou92b,Walls94}.

\section{
Light-to-motion quantum state exchange
}

As depicted in Fig.~2, the two NOPA output fields are assumed to
be incident on separate cavities, each containing a trapped atom 
in the configuration described earlier. Note that 
the output fields from the NOPA are resonant with the respective 
cavity mode frequencies.
We assume that $\Gamma (t)=\Gamma$, a constant, and, for simplicity,
that $\Gamma$ is the same for both configurations.
Denoting the motional mode operators for the two atoms along the
$x$-axis by $\tilde{b}_{1x}$ and $\tilde{b}_{2x}$, respectively, the
two systems are thus described by
\cite{Gardiner93,Carmichael93}
\begin{eqnarray} 
\dot{\tilde{b}}_{jx} &=&
- \Gamma \tilde{b}_{jx} + \sqrt{2\Gamma} \,
\hat{a}_{\rm in}^{(j)}(t) \nonumber
\\
&=& - \Gamma \tilde{b}_{jx} + \sqrt{2\Gamma} \,
\hat{c}_{\rm out}^{(j)}(t-\tau ) \, , \;\;\;\; (j=1,2)
\label{b1eq}
\end{eqnarray}
where $\tau$ is a time delay (assumed the same for both cavities); 
provided the coupling between the NOPA
and the cavities is unidirectional, this delay can essentially
be ignored \cite{Gardiner93,Carmichael93}.

If the bandwidths of the input light fields from the NOPA are
sufficiently broad, in particular if $\kappa_c\gg\Gamma$
[i.e., $\hat{c}_{\rm out}^{(1,2)}(t)$ can be regarded as 
{\em quantum white noise} operators in (\ref{b1eq})], 
then one can perform an average over the input 
fields and derive a master equation (see, e.g., 
\cite{Gardiner91,Walls94}) for the density operator $\rho$ of the 
motional modes alone,
\begin{eqnarray} \label{eq:drhomdt}
\dot{\rho}\, && = 
\Gamma (N+1)(2\tilde{b}_{1x}\rho\tilde{b}_{1x}^\dagger -
\tilde{b}_{1x}^\dagger\tilde{b}_{1x}\rho 
- \rho\tilde{b}_{1x}^\dagger\tilde{b}_{1x}) \nonumber
\\
&& \;\; +\, \Gamma N(2\tilde{b}_{1x}^\dagger\rho\tilde{b}_{1x} - 
\tilde{b}_{1x}\tilde{b}_{1x}^\dagger\rho - 
\rho\tilde{b}_{1x}\tilde{b}_{1x}^\dagger ) \nonumber
\\
&& \;\; +\, \Gamma (N+1)(2\tilde{b}_{2x}\rho\tilde{b}_{2x}^\dagger -
\tilde{b}_{2x}^\dagger\tilde{b}_{2x}\rho 
- \rho\tilde{b}_{2x}^\dagger\tilde{b}_{2x}) \nonumber
\\
&& \;\; +\, \Gamma N(2\tilde{b}_{2x}^\dagger\rho\tilde{b}_{2x} - 
\tilde{b}_{2x}\tilde{b}_{2x}^\dagger\rho - 
\rho\tilde{b}_{2x}\tilde{b}_{2x}^\dagger ) \nonumber
\\
&& \;\; +\, 2\Gamma M(\tilde{b}_{1x}\rho\tilde{b}_{2x}
+ \tilde{b}_{2x}\rho\tilde{b}_{1x}
- \tilde{b}_{1x}\tilde{b}_{2x}\rho -
\rho\tilde{b}_{1x}\tilde{b}_{2x}) \nonumber
\\
&& \;\; +\, 2\Gamma M(\tilde{b}_{1x}^\dagger\rho\tilde{b}_{2x}^\dagger
+ \tilde{b}_{2x}^\dagger\rho\tilde{b}_{1x}^\dagger
- \tilde{b}_{1x}^\dagger\tilde{b}_{2x}^\dagger\rho -
\rho\tilde{b}_{1x}^\dagger\tilde{b}_{2x}^\dagger) \, ,
\end{eqnarray}
with the parameters $N$ and $M$ given, in terms
of the NOPA parameters, by
\begin{equation}
N = \frac{4\epsilon^2\kappa_c^2}{(\kappa_c^2-\epsilon^2)^2} \, ,
\;\;\;\;
M = 2\kappa_c\epsilon\,
\frac{\kappa_c^2+\epsilon^2}{(\kappa_c^2-\epsilon^2)^2} \, .
\end{equation}
This master equation has a steady state solution
\begin{equation}
\rho^{\rm ss} = |\psi_{12}\rangle\langle\psi_{12}| \, ,
\end{equation}
i.e., a pure state, with
\begin{eqnarray} \label{psi12}
|\psi_{12}\rangle &=& S_{12}(r) |0\rangle_{1x} |0\rangle_{2x} 
\nonumber
\\
&=& \left[ \cosh (r) \right]^{-1} \sum_{m=0}^\infty
\left[ -\tanh (r) \right]^m \, |m\rangle_{1x} |m\rangle_{2x} \, ,
\end{eqnarray}
where $|m\rangle_{1x,2x}$ are Fock states of the motional
modes and $S_{12}(r)$ is the two-mode squeezing operator
\cite{Walls94},
\begin{equation}
S_{12}(r) = \exp \left[ r\left( 
\tilde{b}_{1x}\tilde{b}_{2x} -
\tilde{b}_{1x}^\dagger\tilde{b}_{2x}^\dagger 
 \right) \right] \, .
\end{equation}
This operator transforms the mode annihilation operators 
for the atomic motion as
\begin{eqnarray}
S_{12}^\dagger (r) \tilde{b}_{1x} S_{12}(r) &=& 
\cosh (r)\, \tilde{b}_{1x} - \sinh (r)\, \tilde{b}_{2x}^\dagger \, ,
\\
S_{12}^\dagger (r) \tilde{b}_{2x} S_{12}(r) &=& 
\cosh (r)\, \tilde{b}_{2x} - \sinh (r)\, \tilde{b}_{1x}^\dagger \, ,
\end{eqnarray}
where $\cosh (r)=\sqrt{N+1}$ and $\sinh (r)=\sqrt{N}$.
Defining position and momentum operators as
\begin{equation}
Q_j = \tilde{b}_{jx} + \tilde{b}_{jx}^\dagger \, ,
\;\;\;
P_j = -i\left( \tilde{b}_{jx} - \tilde{b}_{jx}^\dagger 
\right) \, ,
\end{equation}
it follows that
\begin{eqnarray}
S_{12}^\dagger (r) \left( Q_1 + Q_2 \right) S_{12}(r) &=& 
e^{-r} \left( Q_1 + Q_2 \right) 
\\
S_{12}^\dagger (r) \left( P_1 - P_2 \right) S_{12}(r) &=& 
e^{-r} \left( P_1 - P_2 \right) \, ,
\end{eqnarray}
and so, in the limit $\epsilon\rightarrow\kappa_c$ 
(i.e., $r\rightarrow\infty$), an EPR state in the positions and
momenta of the two trapped atoms is established. 

The nature of the correlations inherent in the joint state 
(\ref{psi12}) of the atomic motion is most clearly
expressed through the Wigner function for this state
\cite{Ou92b,Walls94}:
\begin{eqnarray}
&& W(q_1,p_1;q_2,p_2) \nonumber
\\
&& \;\;\;\;\;\;\; = \frac{4}{\pi^2} \, 
\exp \left\{ -\left[ (q_1+q_2)^2+
(p_1-p_2)^2\right] e^{+2r} \right\} \nonumber
\\
&& \;\;\;\;\;\;\;\;\;\;\;\;\;\; \times \; 
\exp \left\{ -\left[ (q_1-q_2)^2+
(p_1+p_2)^2\right] e^{-2r} \right\} 
\\
&& \;\;\;\;\;\;\; \rightarrow
C\; \delta (q_1+q_2)\,\delta (p_1-p_2) \;\;\;
{\rm as} \;\; r\rightarrow\infty \; ,
\end{eqnarray}
with $C$ a constant.
This entangled state is achieved in steady state over a time
$t\gg\Gamma^{-1}$. The coupling to the external fields from the
NOPA can then be turned off by setting ${\cal E}_{{\rm L}1,2}$
to zero. The result is a stored EPR state for the motion of
two trapped atoms that would persist for a duration set by 
the timescale for motional decoherence.

\section{
Discussion
}

Before considering some of the interesting possibilities 
offered by this system, we return briefly to some of the 
major assumptions associated with the model. 
Firstly, the finite effect of atomic spontaneous emission
events on the motional state can be estimated as in 
\cite{Parkins99}. This effect can be neglected provided the 
rate of these events is much smaller than the rate $\Gamma$ at 
which the motional steady state is achieved. The condition
one derives by enforcing this inequality amounts to the 
condition of a ``one-dimensional'' atom in cavity QED, 
$C_1=g_0^2/(\kappa_a\gamma )\gg 1$, where $\gamma$ is the 
linewidth (FWHM) of the atomic transition
\cite{Turchette95}.
Note that the experiment of Ref.\cite{Hood98} has achieved
$C_1=70$.
With regards to the trapping potential, harmonic frequencies on 
the order of tens of MHz have been achieved in ion traps,
with corresponding Lamb-Dicke parameters on the order of $0.1$
and smaller \cite{Wineland98}.
Note, however, that large values of the entanglement
parameter $r$ imply population of large-$m$ number states and
a broader spread of the atomic wavepacket. Given that the 
mean excitation number for the state (\ref{psi12}) is 
$\bar{n}=\sinh^2(r)$, a more precise form of the 
Lamb-Dicke assumption would be 
$\eta_x\sqrt{\bar{n}+1}=\eta_x\cosh (r)\ll 1$.
With trap frequencies such as those quoted above, 
the condition $\nu_x\gg\kappa_a$ should be satisfied for
a cavity field decay rate of a few MHz or less; such values
of $\kappa_a$ have not been realized in current experiments
as in Ref.\cite{Hood98}, but potentially could be with
improved cavity finesse as in Ref.\cite{Rempe92}.
Assuming that this is the case, 
likely magnitudes for the rate $\Gamma$ would then
be tens or hundreds of kHz. Finally, timescales for motional
decoherence and heating in recent ion trap experiments are
of the order of milliseconds, with further improvement 
likely \cite{Wineland98}; given the various rates discussed 
above, these effects would not be expected to hamper the 
preparation of the entangled state.

As for applications of this system, further investigation of 
the EPR paradox would obviously be possible, with a variety
of motional state measurements able to be implemented 
on the trapped atoms \cite{Leibfried96,Gardiner97}, 
possibly also via the cavity field \cite{Parkins99}.
In particular, violations of a Bell inequality for the state
(\ref{psi12}) can be obtained 
with measurements that project onto a basis of even and odd
phonon number for each of the trapped atoms 
\cite{Peres93,Banaszek98}.
To the extent that a ``macroscopic'' number of quanta may
in principle be involved, such investigation could also 
address new viewpoints on the compatibility of quantum 
mechanics with local realism \cite{Reid96}.

On a somewhat more applied side is the possibility of using 
the EPR state (\ref{psi12}) for quantum dense coding 
\cite{Bennett92}, or for the {\em teleportation} of the 
quantum state of a system with continuous variables 
\cite{Vaidman94,Braunstein98}, 
generalizing the original discrete-variable teleportation 
protocol of Bennett {\em et al}. \cite{Bennett93}. This 
elegant adaptation of the EPR paradox has in 
fact been realized with light fields, again using optical 
parametric amplifiers and homodyne measurements of quadrature 
amplitudes \cite{Furusawa98}. The scheme outlined in this paper 
opens the door to teleporting an atomic center-of-mass wave
function \cite{Parkins20}, by providing the motional state 
entanglement required by the continuous-variable teleportation 
protocol.

Such a capability is also of considerable interest in the 
related context of quantum computation with trapped 
atoms and light \cite{QC1,QC2}.
Here, we specifically have in mind protocols that combine
quantum information processing with both discrete and
continuous variables \cite{Lloyd99}. Any implementation of a 
qubit (e.g., internal atomic states or photon 
polarization) could be linked with an external degree of 
freedom (e.g., atomic center-of-mass or complex 
amplitude of the electromagnetic field), with the complete 
system viewed as a composite unit (qubit plus {\em qunat} 
\cite{Lloyd99}) for protocols such as quantum communication 
between distant nodes of a quantum network
\cite{Parkins99,Cirac97,vanEnk97,Pellizzari97,vanEnk98}.

Further to this theme, note also that we need not restrict 
ourselves to a single trapped atom at each site. For example,
if there are $K$ atoms inside each cavity, then
focussing the coupling lasers (${\cal E}_{{\rm L}1,2}$) 
sequentially on atoms $1$, $2$, $\ldots$, $K$ at each site
(and neglecting any direct interaction between neighbouring
atoms at each site) would generate a set of pair-wise 
EPR-entangled atoms. 
Alternatively, and perhaps more interestingly, one might 
consider the case in which the $K$ atoms at each site are
{\em simultaneously} coupled to the cavity. Assuming for
simplicity that they have identical coupling strengths, 
then the system would again be described by 
Eq.~(\ref{eq:drhomdt}), but with the replacements 
$\tilde{b}_{1,2x}\rightarrow\tilde{B}_{1,2x}\equiv
K^{-1/2}\sum_{j=1}^K\tilde{b}_{1,2x}^{(j)}$
and $\Gamma\rightarrow K\Gamma$. Such dynamics evidently
leads to a highly entangled state of {\em all 2K atoms},
a situation of potentially great utility and, indeed, of
considerable general interest.

\acknowledgements

ASP gratefully acknowledges support from the Marsden Fund of 
the Royal Society of New Zealand.
HJK is supported by the National Science Foundation, 
by DARPA via the QUIC Institute which is administered by ARO, 
and by the Office of Naval Research.

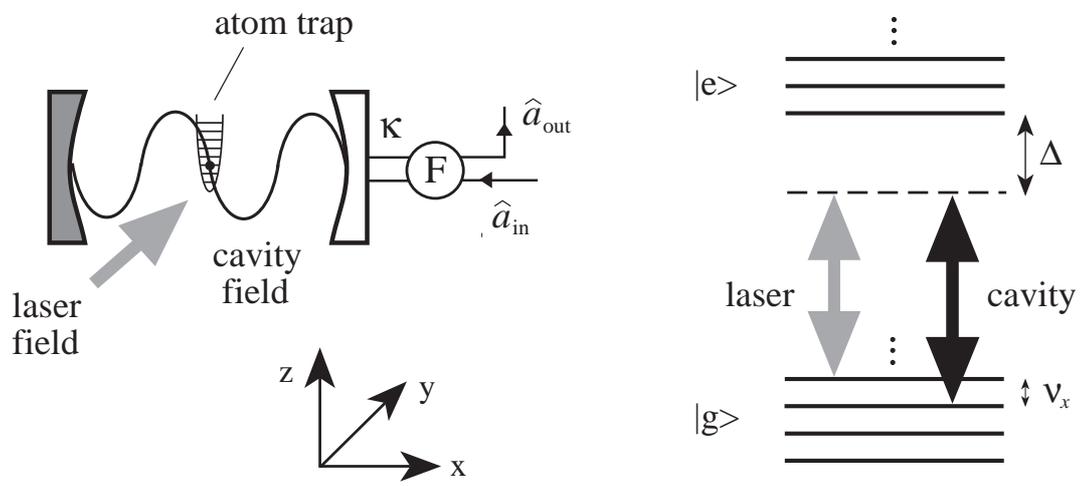
\begin{figure}
\caption{
Schematic of proposed experimental setup and excitation scheme 
for coupling between the motion of a trapped atom and a 
quantized cavity mode of the electromagnetic field, and thence
to a freely propagating external field.
Note that all input and output to the atom-cavity system is assumed
to be through just one mirror; the other mirror is assumed to be 
perfect.}
\end{figure}

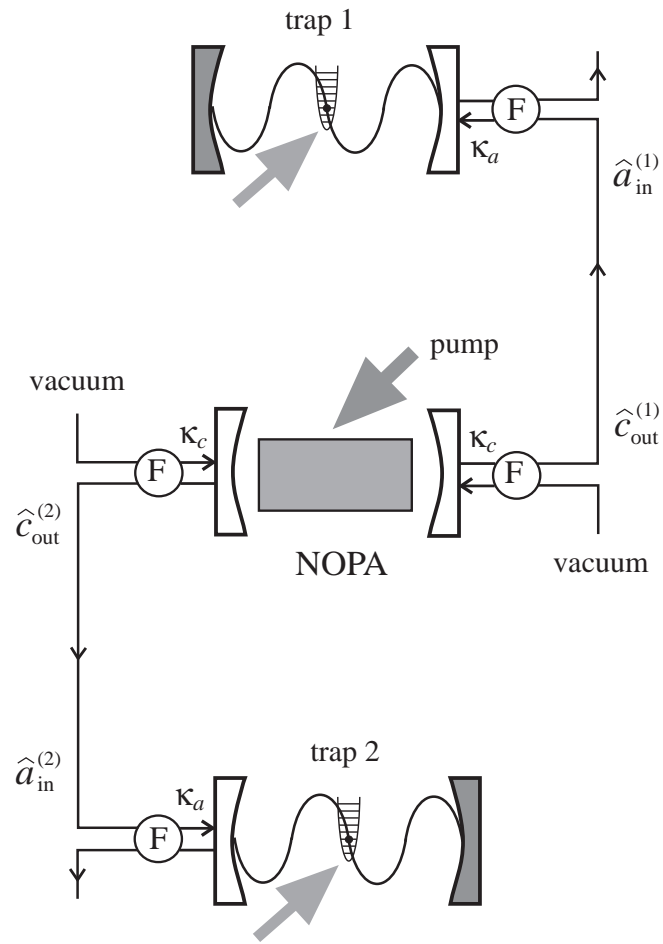
\begin{figure}
\caption{
Preparation of an EPR state of the motion of two separated 
trapped atoms. The output modes from a nondegenerate parametric 
amplifier (NOPA) are incident on two separated 
trapped-atom-cavity configurations of the type depicted in Fig.~1.
Faraday isolators (F) facilitate a unidirectional coupling between 
the entangled light source and the atom-cavity systems.
Note that in practice the outputs from the two NOPA modes might
actually be through the same mirror, but could be separated due
to differing polarizations or frequencies. 
}
\end{figure}

\end{document}